\documentclass[submission,copyright,creativecommons]{eptcs}
\providecommand{\event}{SSV 2012} 
\usepackage{breakurl}             
\usepackage[utf8]{inputenc}
\usepackage{graphicx,thumbpdf}
\usepackage{float}
\usepackage{listings}
\usepackage{multirow}
\usepackage{color}
\definecolor{gray}{rgb}{0.5,0.5,0.5}

\newlength\listinglength
\title{CTGEN - a  Unit Test Generator for C}
\author{Tatiana Mangels\thanks{The author has been supported by Siemens AG through a research grant of the Graduate School on Embedded Systems GESy at the University of Bremen ({\tt http://www.informatik.uni-bremen.de/gesy})}
\institute{University of Bremen\\
Germany}
\email{tatiana@informatik.uni-bremen.de}
\and
Jan Peleska\thanks{The author's research is funded by the EU FP7 COMPASS project under grant agreement no.287829}
\institute{University of Bremen\\
Germany}
\email{jp@informatik.uni-bremen.de}
}
\def\titlerunning{CTGEN - a Unit Test Generator for C}
\def\authorrunning{T. Mangels and J. Peleska}
\begin{document}
\maketitle


\begin{abstract}
We present a new unit test generator for C code, CTGEN. It generates test data for C1 structural coverage 
and functional coverage based on pre-/post-condition specifications or internal  assertions. 
The generator supports automated stub generation, and
data to be returned by the stub to the unit under test (UUT) may be
specified by means of constraints. The typical   
application field for CTGEN is embedded systems testing; therefore the tool can cope with the typical
aliasing problems present in low-level C, including pointer arithmetics, structures and unions.
CTGEN creates complete test procedures which are ready to be compiled and run against the UUT.
In this paper we describe the main features of CTGEN, their technical realisation, and we elaborate
on its performance in comparison to a list of competing test generation tools. Since 2011, CTGEN is used 
in industrial scale test campaigns for embedded systems code in the automotive domain.

\end{abstract}


\section{Introduction}

{\it Unit testing} is a well-known method, widely used in practice, by
which a single program function or method (hereafter referred to as {\it
  module} or the {\it unit under test (UUT)})   is tested separately
with respect to its functional correctness\footnote{In this paper we
  disregard tests investigating non-functional properties, such as
  worst case execution time or  
  absence of run-time errors, since these are often more successfully
  investigated by means of formal 
  verification, static analysis or abstract interpretation.}. 
  The test data for performing a module test specifies initial values
  for input parameters, global variables, and for the data to be set
  and returned by sub-functions called by the UUT. The test results 
  are typically checked by means of pre-/post-conditions, that is,
  logical conditions relating the  
  program pre-state to its post state after the module's
  execution. More complex correctness conditions, such as the number
  of sub-function calls performed by the UUT may also be specified by
  means of pre-/post-conditions, if auxiliary variables are
  introduced, such as counters for the number of sub-function calls  
  performed.
  The manual elaboration of test data and the development of test
  procedures exercising these data on the UUT is time consuming and 
expensive. 

{\it Symbolic execution} \cite{king, clarke} is a well-known technique that
addresses the problem of automatic generation of test inputs. 
Symbolic execution is similar to a normal execution process, the
difference being that the values of program inputs are symbolic
variables, not concrete values. When a variable is updated to a new
value it is possible, that this new value is an expression over such symbolic
variables. When the program flow comes to a branch, where the
condition depends on a symbolic variable, this condition can be
evaluated both to {\it true} or {\it false}, depending on the value of
a symbolic variable. Through the
symbolic execution of a path, it becomes a {\it path condition}, which
is a conjunction of all branch conditions occurring on the path. 
To reason about path conditions and hence about feasibility of
paths a constraint solver is used. When the solver determines a
path condition as feasible, it calculates concrete values
which can then be used as concrete inputs to explore the corresponding
path.

In this paper we present CTGEN, an automatic test generation tool,
based on symbolic execution. The objective of CTGEN
is to cover every branch in the program, which is an
undecidable problem, so in practice CTGEN tries to generate a test
that produces as high a coverage for the module under tests as  
possible. For each unit under test CTGEN performs symbolic analysis
and generates a test in RT-Tester syntax \cite{rtt_manual}, which can
be directly compiled and executed. For the 
definition  of expected behaviour of the UUT CTGEN
provides an annotation language, which also offers support for referencing
functional requirements for the purpose of traceability from tests to requirements.
Apart from atomic integral data types,   CTGEN supports floating point variables,
pointer arithmetics, structures and arrays and can cope with the typical aliasing problems
in C, caused by array and pointer utilisation. Recursive functions, dynamic 
memory, complex dynamic data structures with pointers (lists, stacks etc.) and
concurrent program threads are not supported. CTGEN does
not check the  module
under test for run-time errors, but rather  
delegates this task to an abstract interpreter
which is developed in our research group~\cite{sonolar}.

CTGEN does not rely on knowledge about all parts of the program (such as
undefined or library functions). Where several other unit test
automation tools \cite{cute, exe, dart} 
fall back to the invocation of the original sub-function code with concrete inputs if an external function occurs on the explored path,
CTGEN automatically
generates a {\it mock object}   replacing the external function by a 
test stub with the same signature. Furthermore it
calculates values for the stub's return data, output parameters and global
variables which can be modified by the stubbed function in order to
fulfil   a path condition.  In this way CTGEN can also simulate 
 exceptional behaviour of external functions. 
It is possible but
not required to
customise stub behaviour by using pre- and post-conditions described
in Section \ref{annotation_language}. If no restrictions were made, however,
the stub's actions can deviate from the real behaviour of the external
function. 

\paragraph{Main contributions.}
CTGEN is distinguished from other unit test automation tools    
by the following capabilities: (1) handling of external function calls by means of mock objects, 
(2) symbolic pointer and offset handling, and (3) requirement tracing.
As  pointed out in \cite{concolic:study}, pointers and external
function calls are the most important challenges for test data
generation tools. CTGEN handles both as described in this paper. 
Furthermore we provide heuristics for the selection of paths through
the UUT in such a way, that the coverage may be achieved
with as few test cases as possible.

\paragraph{Overview.} 
This paper is organised as follows. Section \ref{architecture}
presents an overview of the CTGEN architecture. Section
\ref{annotation_language} introduces an annotation language developed
to define expected behaviour of the UUT and to support requirements tracing. 
In Section \ref{stcg} we discuss
test case generation strategies used by CTGEN. Section
\ref{symbolic_execution} describes the algorithm  
for handling pointer comparisons and pointer arithmetics. We
show experimental results and compare CTGEN to other test data
generation tools in Section \ref{experimental_results}, discuss
related work in Section \ref{related_work} and present a conclusion in Section
\ref{conclusion}. 
 
\section{Architecture}\label{architecture}

\begin{figure}
\begin{center}  
\includegraphics[angle=270, scale=0.6]{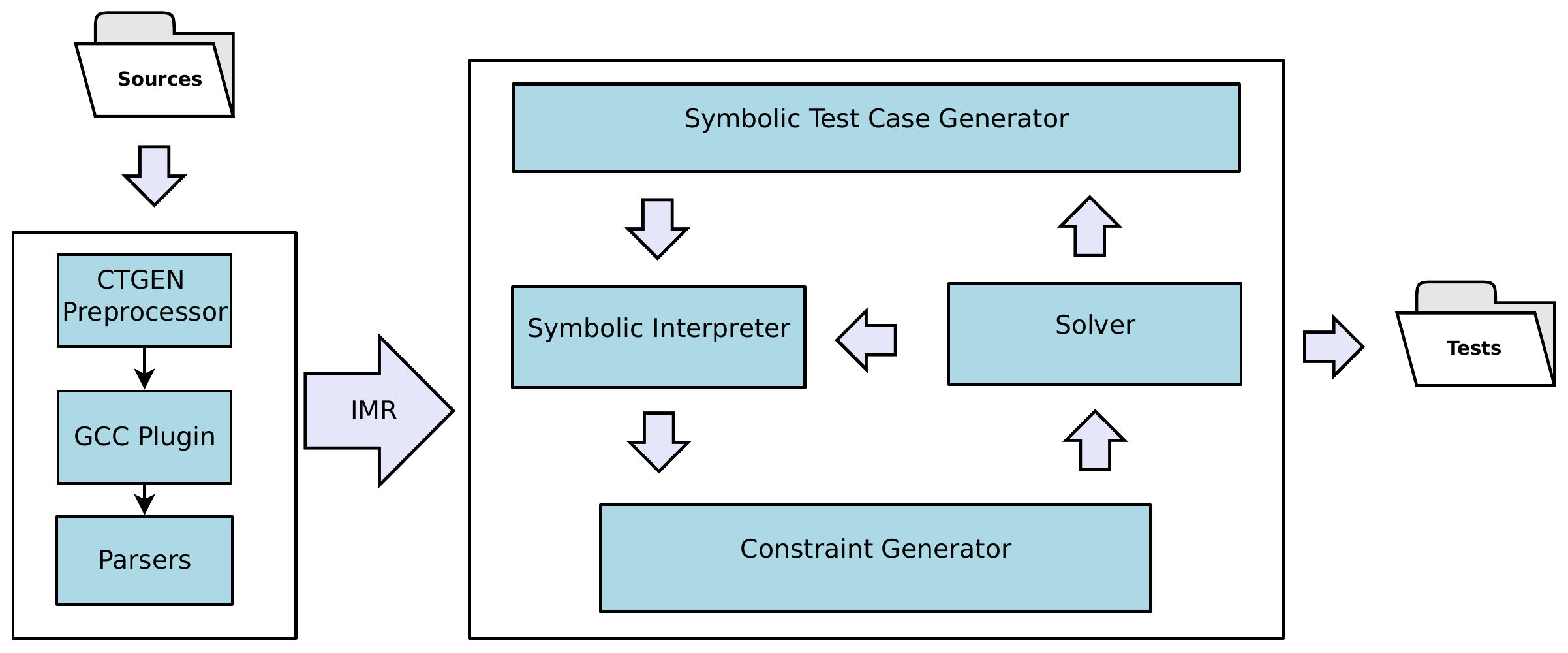}
\caption{CTGEN overview.}
\label{pic:overview}
\end{center}  
\end{figure}

CTGEN is structured into two main components  (see Fig.~\ref{pic:overview}):

The {\it preprocessor} 
operates on the UUT code. It consists
  of (1) the CTGEN preprocessor transforming  code
    annotations as described in Section~\ref{annotation_language}, (2) a
  \verb+GCC plugin+ based on \cite{loeding}, compiling
  the prepared source code into a textual specification consisting of
    Control Flow Graphs (CFGs) in 3-address code and symbol table
  information like function signatures, types and variables, and (3)
  parsers, transforming  CFGs and symbol table information into the
  Intermediate Model Representation (IMR).

The {\it analyser}  operates on the IMR, its building
  blocks and their interactions are
  described below. The {\it Symbolic Test Case Generator} is responsible
  for lazy expansion of the CFGs related to  the function under test and its sub-functions. Moreover, it handles the  selection of paths, each beginning with the start node of the CFG, and
  containing yet uncovered transitions (for more details see Section~\ref{stcg}). If such path
  can be found, it is passed to the {\it Symbolic Interpreter}, which
  traverses the path and symbolically calculates the effect of its statements in
    the memory model.
    As soon as the next node on the path is guarded by
   a non-trivial   condition, the {\it Constraint Generator}
   \cite{peleska} is called and resolves all pointers and array
   references occurring in   this condition. It also passes the resulting constraint to
   the {\it Solver}. CTGEN uses a  SMT solver, SONOLAR, which has been developed in
   our research group \cite{sonolar} and supports integral and
   floating point datatypes, arrays and bit vectors. 
   If the solver can find a solution
   for the   constraint, the solution is passed  back to the Symbolic
     Interpreter, which continues to follow the path under
   investigation. 
   Otherwise, if the constraint is infeasible or the
   path is completed, the solver passes the result   to the {\it
     Symbolic Test Case Generator}. It learns from infeasibility and tries to
   produce another path containing still uncovered transitions, or
   tries to continue the completed path. When
   no such paths can be found, a unit test is generated based on the
   collected solutions (if any)  and is stored in the file system. 


\section{Annotation Language}\label{annotation_language}

Annotations in CTGEN allow   users to specify the expected behaviour of
functions. For the definition of annotations we have chosen the approach
used in sixgill \cite{sixgill}: they are specified as GCC
macros which are understood by CTGEN. Thus the annotations can be turned on and
off as needed. The arguments of the annotations follow standard C
syntax, so that no additional expertise is required for the user. All
annotations are optional -- if there are none, CTGEN will try to cover
all branches and detect unreachable code, using arbitrary type-compatible input data. 

Pre- and post-conditions are  defined as follows:

\lstset{language=C++, frame=single, captionpos=b,
basicstyle=\scriptsize, emph={__rtt_precondition,
  __rtt_postcondition, __rtt_testcase, __rtt_aux, __rtt_initial,
  __rtt_assert, __rtt_assign,
  __rtt_return, __rtt_modifies, @rttAssert, @rttCall},
emphstyle=\textbf, breaklines=true}
\noindent
\begin{minipage}{\textwidth}
\begin{lstlisting}
__rtt_precondition(PRE);
__rtt_postcondition(POST);
\end{lstlisting}
\end{minipage}

A precondition indicates, that the expected behaviour of the specified
function is only guaranteed if the condition \verb+PRE+ is true. A
post-condition specifies, that after the execution of a function
the condition \verb+POST+ must hold. 
Pre- and post-conditions have to be defined directly at 
the beginning of a function body. {\tt PRE} and {\tt POST} 
are   boolean C expressions, including function calls. 
All variables occurring in these conditions must be global, input or
output parameters, or refer to the return value of the specified
function. 
To specify conditions involving the return value of the UUT
the   CTGEN variable \verb+__rtt_return+ may be used.
Annotation \verb+__rtt_initial(VARNAME)+ can be used within other annotation expressions -- in particular, in post-conditions  -- for referring to the initial   value of the variable \verb+VARNAME+, valid before the function was executed.

To reason over local variables, auxiliary variables are used. Auxiliary variables 
can never occur in assignments to non-auxiliary variables or in control flow
conditions \cite{apt, peleska}.  They can be defined as
follows: 
\begin{lstlisting}
__rtt_aux(TYPE, VARNAME);
\end{lstlisting}
In this way an auxiliary
variable of type \verb+TYPE+ with the name  \verb+VARNAME+ will be declared
and can be used in the following CTGEN annotations in the same way as
regular variables. 

For a more detailed specification of the expected behaviour of the
function, test cases can be used:
\begin{lstlisting}
__rtt_testcase(PRE, POST, REQ)
\end{lstlisting}
The argument \verb+PRE+ defines a pre-condition  and the argument \verb+POST+ a
post-condition of the current testcase. 
The argument \verb+REQ+ is a string tag defining  a functional
requirement that corresponds to the pre- and post-condition of this
test case. If there is more than one requirement, they can be listed
in a comma-separated list. 

%
%
%
Users can specify global variables which are allowed for modification
in the current function by means of  annotation:  
\begin{lstlisting}
__rtt_modifies(VARNAME)                
\end{lstlisting}
CTGEN traces violations,
even in cases where a prohibited variable is modified by means of pointer
dereferencing. For each breach of a modification rule an assertion is generated, which
records the line number where the illegal modification occurred, e.~g.
\begin{lstlisting}
// violated var VARNAME in line(s) 1212, 1284
@rttAssert(FALSE);
\end{lstlisting}

The \verb+__rtt_assign(ASSIGNMENT)+ annotation is intended for assignments to auxiliary
variables. In the following example an auxiliary variable \verb+a_aux+ is first declared using
\verb+__rtt_aux()+; it may then be used in a post-condition's
expression. To define its value, \verb+__rtt_assign()+ is used in the function body.

\noindent
\begin{minipage}{\textwidth}
\begin{lstlisting}
__rtt_aux(int, a_aux);
__rtt_postcondition(a_aux == 0);
...
int b;
...
__rtt_assign(a_aux = b);             
\end{lstlisting} 
\end{minipage}

\verb+__rtt_assert(COND)+ can be used in different places of the
function to ensure a specific property. If   condition \verb+COND+ is seen to fail during test generation
an assertion recording the line number where the violation occurs is inserted
into the generated test.

An example of a specification of the expected behaviour of a function is
illustrated in Fig.~\ref{pic:precond}. The function \verb+alloc()+ returns a
pointer \verb+allocp+ to \verb+n+ successive characters if there is
still enough room in the buffer \verb+allocbuf+ and zero if this is
not the case. First, by using \verb+__rtt_modifies+ we state that
\verb+alloc()+ can only modify \verb+allocp+, and modification of \verb+allocbuf+ is
consequently prohibited. Annotation \verb+__rtt_precondition+ specifies
that the expected behaviour of \verb+alloc()+ is guaranteed only if
the parameter \verb+n+ is greater or equal zero and \verb+allocp+ is
not a \verb+NULL+-pointer. Furthermore the condition \verb+__rtt_postcondition+ states
that after the execution of the function under test \verb+allocp+
must still be within the bounds of the array \verb+allocbuf+. Finally, test cases
are defined for the situations where (a) memory can still be allocated
and (b) not enough memory is available.

\begin{figure}[H]
\lstset{language=C++, frame=single, captionpos=b,
basicstyle=\scriptsize, emph={__rtt_precondition,
  __rtt_postcondition, __rtt_testcase, __rtt_aux, __rtt_initial,
  __rtt_return, __rtt_modifies},
emphstyle=\textbf, breaklines=true, numbers=left, numberstyle=\tiny\color{gray},
numbersep=5pt}
\begin{lstlisting}
char allocbuf[ALLOCSIZE];
char *allocp = allocbuf;

char *alloc(int n){
  __rtt_modifies(allocp);
  __rtt_precondition(n >= 0 && allocp != 0);
  __rtt_postcondition(allocp != 0 && allocp <= allocbuf + ALLOCSIZE);
  __rtt_testcase(allocbuf + ALLOCSIZE - __rtt_initial(allocp) < n, 
                 __rtt_return == 0, 
                 "CTGEN_001");
  __rtt_testcase(allocbuf + ALLOCSIZE - __rtt_initial(allocp) >= n, 
                 __rtt_return == __rtt_initial(allocp), 
                 "CTGEN_002");

  char *retval = 0;
  if(allocbuf + ALLOCSIZE - allocp >= n){
    allocp += n;
    retval = allocp - n;
  } 
  
  return retval;
}
\end{lstlisting}
\caption{Example: Specification of expected behaviour.} 
\label{pic:precond}
\end{figure}


\section{Symbolic Test Case Generation}\label{stcg}

As described in Section~\ref{architecture}, the symbolic test case
generator is responsible for the selection of test cases, which the
symbolic interpreter then tries to cover. As a criteria to reason
about completeness of testing we use code coverage. CTGEN supports the
following two  coverage criteria:  
(1) {\it statement coverage} (C0), which requires, that each statement
in the program is executed at least once, and
(2) {\it decision coverage} (C1), which requires that additionally to statement
coverage each decision in the program was evaluated at least once to
{\it true} and at least once to {\it false}. 


A central data structure
in the symbolic test case generator is a {\it Symbolic Test Case Tree}
(STCT), which
 stores bounded paths through
the control flow graph of the UUT. To build a
STCT the CFG is expanded node by node. During expansion of CFG node $n$,
each outgoing edge  of $n$ is analysed and each of its target nodes 
becomes a new corresponding STCT leaf, even if this target node
already has been expanded. The nodes are labeled 
with a number $k$, so that $(n, k)$ is a
unique identifier of a STCT node, while $n$ may occur several times in the STCT
if it lies on a cyclic CFG path. The STCT root corresponds to the CFG start node (see  
 \cite{badban} for further details).

When it is required to select a new test case, the symbolic test case
generator takes an edge in the CFG, which is still uncovered. Subsequently
it finds a
corresponding STCT edge (since nodes $n,n'$ may occur several times in
the STCT,   edges $n\rightarrow n'$ may occur several times as well)
and traces it bottom-up to the start 
node. This trace is then returned to the test data
generator by the symbolic test case generator for further investigation.

The depth-first search used by several test
generating tools \cite{pathcrawler:experience, cute, dart, exe}
allows to reuse the information of the shared part of the execution trace,
but on the other hand can cause
the generator to get ``stuck'' analysing a small part of the program and
generating a lot of new test cases but no (or little) new
coverage. Pex \cite{pex} avoids using depth-first and  backtracking techniques by storing
the information of all previously executed paths. 
CTGEN behaves similarly: during the symbolic execution of
a trace it stores information already gained   in the form
of  computation histories and  path constraints associated with 
 each branch point of the trace under consideration.

\begin{figure}
\begin{center}  
\includegraphics[width=0.96\textwidth]{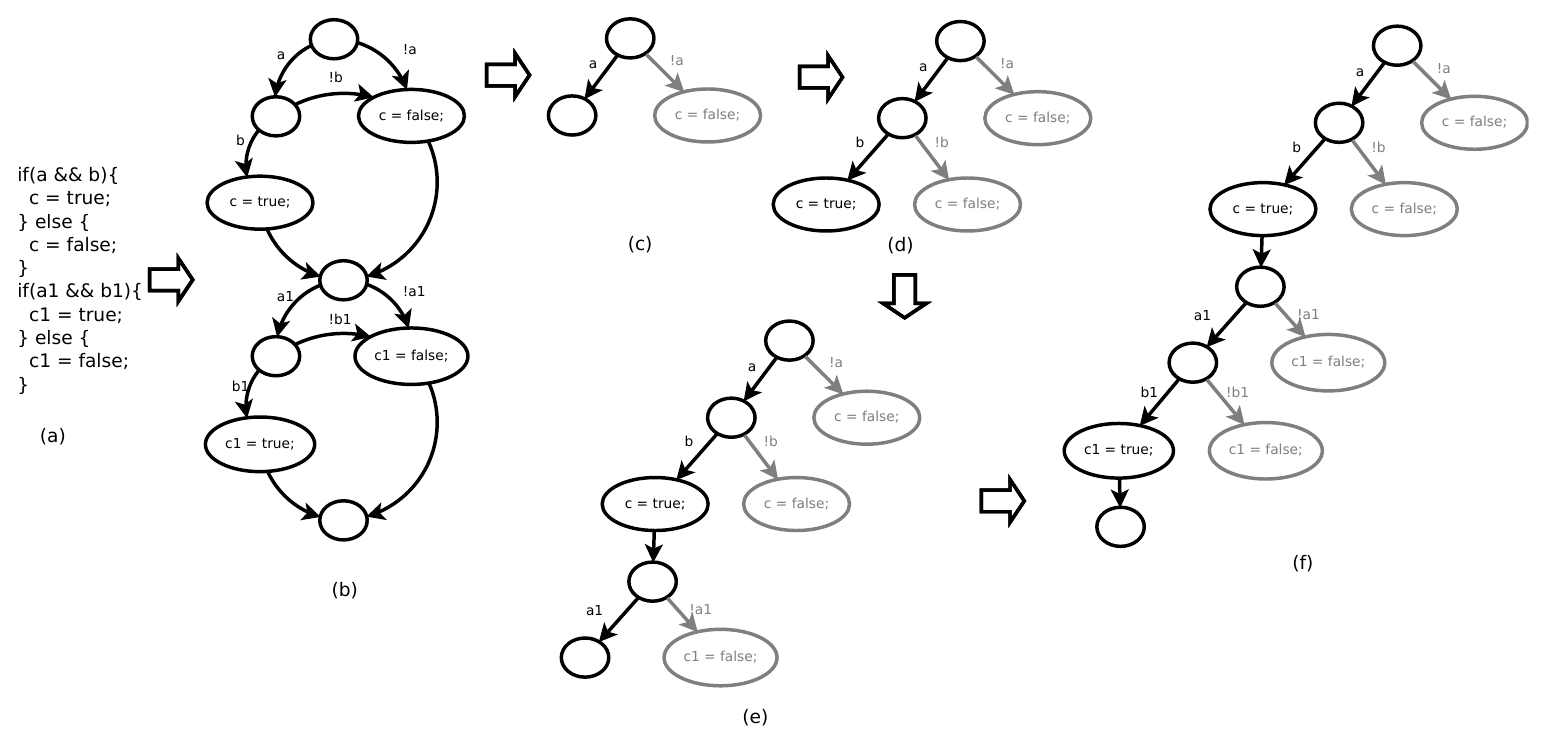}
\caption{Expansion/Selection Example.}
\label{pic:cfg_stct}
\end{center}  
\end{figure}

Our expansion and selection strategies are motivated as follows: 
(a) The larger the explored range of the variable values
  participating in the guard conditions closest to the start node, the
  higher the probability to cover more branches depending on these
  variables further down the CFG. 
  So we prioritise edges after their proximity to the start node.
  This allows achieving more coverage when it is not possible to explore
  the function completely due to its size. Furthermore this approach
  minimises the number of paths that must be explored to achieve
  100 \% C1 coverage, which in turn reduces the overall time for test generation.
(b) A path is interesting for the test case generation for
  C1 coverage only if it contains still uncovered edges with 
  non-trivial guard conditions, since otherwise no new coverage can be
  achieved by interpreting this path. So we expand until the STCT
  contains a new uncovered edge whose guard condition 
  is not always {\it true}, or until   no
  expansion can find any additional uncovered edges.  At the same time we try to
  minimise the size of a STCT, so we stop the expansion process as
  soon as an uncovered edge with non-trivial guard condition
  occurs. We call this approach {\it incremental expansion}. 
(c) To further minimise the size of a STCT we incrementally
  expand  only the end node of the 
  path under  consideration (initially the root
  node), select the continuation for it and hand it over to the solver. We
  continue this step by step until the path is
  complete. After that and according to prioritisation of edges, a new path  is
  selected. If the selected path is infeasible, the
  responsible branch is deleted from the STCT, and the selection/expansion
  process is continued with the alternative branch.

The loop constructs are unwound according to our
expansion strategy: the loop body is incrementally expanded until 
the exit condition is reached. If the exit condition can be
evaluated to {\it true}, the loop is exited. Otherwise it is further
expanded. This process is bounded by a configurable command line
parameter, that 
defines the maximal possible depth of the STCT.
 
A simple example of our expansion/selection strategy is illustrated in
Fig. \ref{pic:cfg_stct}. Fig. \ref{pic:cfg_stct}(a) shows a program,
its corresponding CFG is presented in
Fig. \ref{pic:cfg_stct}(b). After the initial
expansion (Fig. \ref{pic:cfg_stct}(c)) a trace consisting of
edge $<a>$ is selected (nodes
and edges, that belong to the trace are drawn black, those that do not
belong are drawn grey). After the trace is interpreted and evaluated as
feasible, its last node is expanded (Fig. \ref{pic:cfg_stct}(d))
and a new trace (the continuation of the last one) consisting of egdes
$<a>$ and $<b>$ is
selected. After this trace is interpreted as well, its last node $(c=true;)$
is expanded until the new edge whose guard condition is not always {\it
  true} appears
(Fig. \ref{pic:cfg_stct}(e)). Afterwards a new trace (continuation of the
previous one) is selected. This process continues until the path is complete
(Fig. \ref{pic:cfg_stct}(f)). Finally a new trace, corresponding to the
prioritisation of edges, will be selected  (here $<!a>$) and the whole
process will be repeated. 


\section{Symbolic Execution}\label{symbolic_execution}

Due to aliasing in C/C++ the value of a variable can be changed not
only by a direct reference to its  name, but also by assignments to 
de-referenced pointers pointing to this specific variable. 
In the case of arrays different index expressions may reference the same array element.
This makes it
difficult to  identify variable changes along program paths involving pointer and array expressions.
To solve this problem, a memory model consisting of  a history of {\it memory
items} was introduced in \cite{peleska, loeding}. 
Each memory item is defined by its base address, offset, length,
value expression and time interval (measured in computation steps) where it is valid. Computations are
defined as memory modifications which are stored in corresponding
memory items. Furthermore, the stored values are not resolved
to concrete or abstract valuations, but are specified
symbolically. This approach allows not only to find the actual
memory area where a new value is written to,
but also enables us to handle  pointer comparisons and pointer
arithmetics as described below.


\lstset{language=C++, captionpos=b,
basicstyle=\tiny, 
emphstyle=\textbf, breaklines=true, frame=none}
\begin{table}
\begin{tabular}{|c |c |c |c|}
\hline
C code & path constraint & solver solution & generated test case \\
\hline
\begin{lstlisting}
void test(char *p1, 
          char *p2){
  if(p1 < p2){
    ERROR;
  }
}
\end{lstlisting}
&
\begin{lstlisting}
p1@baseAddress == p2@baseAddress &&
p1@offset  <  p2@offset &&
0 <= p1@offset < 10 &&
0 <= p2@offset < 10
\end{lstlisting}
&
\begin{lstlisting}
p1@baseAddress = 2147483648
p2@baseAddress = 2147483648
p1@offset = 0
p2@offset = 7
\end{lstlisting}
&
\begin{lstlisting}
char* p1, p2;
char p1__autogen_array[10];
unsigned int p1__autogen_offset;
unsigned int p2__autogen_offset;

p1 = p1__autogen_array;
p2 = p1__autogen_array;

p1__autogen_offset = 0;
p1 += p1__autogen_offset;
p2__autogen_offset = 7;
p2 += p2__autogen_offset;

@rttCall(test(p1, p2));
\end{lstlisting}

\\
\hline
\end{tabular}
\caption{Pointer handling example.}
\label{tab:ptr_example}
\end{table}

\paragraph{Pointer Handling}
We handle all memory areas pointed to by   pointers as arrays with configurable size. By  
abstracting pointers to integers we achieve that constraints over pointers
can be  solved by a solver capable of integer arithmetics. 
Each pointer $p$ is associated with a
pair of unsigned integers $(A, x)$, where $A$ corresponds to a base address of the memory
area $p$ points into  and
$x$ is its offset ($p = A + x$). Whenever two pointers $p_i = (A_i,x_i), \; i = 1,2$ occur in
an expression  $(p_1 \; \omega \; p_2)$ with some comparison operator $\omega$, we construct the following
constraint: 
$$A_1 == A_2 \; \&\& \; x_1 \; \omega \; x_2 \; \&\& \; 0 \le x_1
< dim(p_1) \; \&\& \; 0 \le x_2 < dim(p_2) \; \&\& \;
dim(p_1) == dim(p_2). $$
where $A_1 == A_2$ ensures, that $p_1$ and $p_2$ point to members of the
same memory portion, $x_1 \; \omega \; x_2$ reflects the pointer expression and $0 \le x_i
< dim(p_i)$ guarantees, that the pointer stays within array bounds.
To illustrate how CTGEN handles pointer operations we use the simple
example shown in Table~\ref{tab:ptr_example}. Consider function
\verb+test()+: to reach the line with an error, input pointers \verb+p1+ and
\verb+p2+ should fulfil   \verb+p1 < p2+. The symbolic
interpreter generates a path constraint according to our approach
with the size for the auxiliary arrays configured by the user over the
command line argument equal to 10 (if it is not configured, default
size will be taken), because this memory size 
makes condition \verb+p1 < p2+ feasible (any size $\geq 2$  would suffice).
All auxiliary variables used here are of type \verb+unsigned int+ so that
the solver can easily solve the generated path constraint. CTGEN
considers the calculated
base address as a unique identifier of an auxiliary array: so, if
the identifier appears for the first time, a new array is created, if
the identifier is already known, the corresponding auxiliary array is
taken. In our example identifier 2147483648 appears for the first time
in the solution for \verb+p1@baseAddress+, so the new array
\verb+p1__autogen_array+ is created and the pointer \verb+p1+ is
initialised with it. When the same identifier appears in the solution for
\verb+p2@baseAddress+, it is already known to CTGEN, so \verb+p2+ is
also initialised with \verb+p1__autogen_array+. Then offset values
are processed and pointers are  modified accordingly. With this test input
the erroneous code in  \verb+test()+ is uncovered.

\begin{table}
\begin{tabular*}{\textwidth}{@{\extracolsep{\fill}} |c|c|c|}
\hline
C code & path constraint & solver solution \\
\hline
\begin{lstlisting}
void test(int p1, int p2){
  __rtt_modifies(globalVar);
  globalVar = -p2;
  if(func_ext(p1) > p2 && func_ext(p2) == p1 && 
     globalVar == p2){
    ERROR;
  }
}
\end{lstlisting}
&
\begin{lstlisting}
func_ext@RETURN@0 > p2 &&
func_ext@RETURN@1 == p1 &&
globalVar@func_ext@1 == p2
\end{lstlisting}
&
\begin{lstlisting}
func_ext@RETURN@0 = 21
func_ext@RETURN@1 = 0
globalVar@func_ext@1 = -1
p1 = 0
p2 = -1
\end{lstlisting}
\\
\end{tabular*}

\begin{tabular*}{\textwidth}{@{\extracolsep{\fill}} | c | c |}
\hline
generated test case & generated stub \\
\hline
\begin{lstlisting}
extern unsigned int func_ext_STUB_testCaseNr;
extern unsigned int func_ext_STUB_retID;
extern int func_ext_STUB_retVal[2];

int p1, p2;

/***** STUB func_ext *****/
func_ext_STUB_testCaseNr = 0;
func_ext_STUB_retID = 0;

/* set values for return */
func_ext_STUB_retVal[0] = 21;
func_ext_STUB_retVal[1] = 0;
/***** end STUB func_ext *****/

p1 = 0;
p2 = -1;
@rttCall(test(p1, p2));
\end{lstlisting}
&
\begin{lstlisting}
int func_ext(int a){
  @GLOBAL:
    unsigned int func_ext_STUB_testCaseNr;
    unsigned int func_ext_STUB_retID;
    int func_ext_STUB_retVal[2];
  @BODY:
    func_ext_RETURN = func_ext_STUB_retVal[func_ext_STUB_retID%2];
    if(func_ext_STUB_testCaseNr == 0){
      if(func_ext_STUB_retID == 1){
	globalVar = -1;
      }
    }
    func_ext_STUB_retID++;
};
\end{lstlisting}
\\
\hline
\end{tabular*}
\caption{External function handling example.}
\label{tab:stub_example}
\end{table}

\paragraph{External Function Handling}
When an external function call appears on the path under
consideration, the return value of this external function, its output parameters
and all global variables allowed for modification are handled as
symbolic {\it stub variables}. These symbolic stub variables can  possibly be modified by
this call. A stub variable holds the information about the stub function to which
it belongs and if it corresponds to the return value, the output parameter or
global variable, changed by this stub. 

To illustrate how this algorithm works we
use the example shown in Table \ref{tab:stub_example}. In function
\verb+test()+ the external function \verb+func_ext()+ is called twice. To
reach the line with an error, \verb+func_ext()+  must return a value
that is greater than the value of the parameter \verb+p2+ by the
first call. Furthermore, by the second call it must return a value
that is equal  to the value of the parameter 
\verb+p1+. The symbolic interpreter analyses, what could possibly be 
altered by \verb+func_ext()+ and creates the stub 
variables \verb+func_ext@RETURN+ and
\verb+globalVar@func_ext+. The constraint generator generates a path 
constraint as shown in Table \ref{tab:stub_example}. The occurences of
the stub variables are versioned (actually, in CTGEN all variables in path
constraint are versioned \cite{peleska}, but this is ignored here to keep the example
simple). The version of a stub variable corresponds to the
running number of the call of the external function. Here
\verb+func_ext@RETURN@0+ corresponds to the return value of the first call and
\verb+func_ext@RETURN@1+ to the return value of the second one. 
Now consider the generated test case and the generated stub in Table
\ref{tab:stub_example}. 
As was already mentioned, CTGEN generates tests in RT-Tester syntax. 
An array \verb+func_ext_STUB_retVal+ of
size two (corresponding to the number of calls of the function
\verb+func_ext()+) is created to hold the calculated return
values. These values are stored by the test driver according to their
version. The variable \verb+func_ext_STUB_retID+ corresponds to the
running number of the stub call. It is reset by the test driver before each call of UUT
and incremented by the corresponding stub each time it is
called. Since one test driver can hold many test cases, the variable
\verb+func_ext_STUB_testCaseNr+ that corresponds to the 
number of test case is created. This variable is set by the test driver. The
value of  the global variable \verb+globalVar+ is set by the stub if the number
of the stub call and the test case number match the calculated ones for
this global variable.

A detailed description how to perform
symbolic execution in presence of arrays and primitive datatypes can
be found in \cite{peleska}. 


\section{Experimental Results}\label{experimental_results}

The experimental evaluation of CTGEN and comparison with competing tools 
was performed both with synthetic examples evaluating the tools' 
specific pointer handling capabilities and with embedded systems code
from an industrial automotive application. The latter presented specific challenges: 
(1) the code was automatically generated from Simulink models. This made automated 
 testing mandatory since   small model changes  considerably affected the structure 
of the generated code, so that re-use of existing unit tests was impossible 
if models had been changed. (2) Some units were exceptionally long 
 because insufficient HW resources required 
to reduce the amount of function calls.

\begin{table}
\scalebox{0.9}{
\begin{tabular}{r | r r r r r r}
      & Executable Lines & Branches & Time & Nr of Test Cases & Lines Coverage & Branch Coverage \\
\hline
$f_1()$ & 714 & 492 & 31m27.098s & 59 & 95,1 \% & 89,0 \% \\
$f_2()$ & 50 & 30 & 0m1.444s & 8 & 100 \% & 100 \% \\
$f_3()$ & 11 & 4 & 0m0.228s & 3 & 100 \% & 100 \% \\
\end{tabular}
}
\caption{Experimental results on some functions of HELLA
  software.}
\label{tab:results}
\end{table}

Table~\ref{tab:results} shows the results achieved by CTGEN in
the automotive test project on some selected functions. The most challenging
function was $f_1$ with over 2000 lines of
code (714 executable lines), using structures, bit vectors, pointer parameters and
complex branch conditions. Nevertheless CTGEN was able to generate
95,1\% line and 89,0\% branch coverage with 59 automatically generated test
cases. 
Furthermore, by using
preconditions as guides for CTGEN to cover parts of code further
down in the CFG, it was possible to increase the coverage even more.
Function $f_2$ with 50 executable lines of code (about 300 lines of
code) represents a typical function in the project. For such functions
CTGEN achieved 100\% C1 coverage.
Function $f_3$ includes pointer comparison, pointer dereferencing and
a \verb+for+-loop with an input parameter as a limit. However, due to small
branching factor CTGEN achieves 100\% coverage with only 3 test cases
and a generation time under one second.
Summarising, CTGEN proved to be efficient for industrial test campaigns in the embedded domain
and considerably reduced the over all project efforts.

\lstset{language=C++, captionpos=b,
basicstyle=\tiny, 
emphstyle=\textbf, breaklines=true, frame=none}
\begin{table}
\scriptsize
\begin{tabular}{|r | c | c | c | c |}
\hline
  & CTGEN & \multicolumn{2}{c|}{KLEE} & PathCrawler \\
\hline
$f_1()$ (714 lines, 492 branches )& & \multicolumn{2}{c|}{} &  \\
\hline
Time & 31m27.098s & 16m50s & 586m16.590s & - \\
Nr of Test Cases & 59 & 1120 & 24311 & - \\
Lines Coverage &  95,1 \% & 77,9 \% & 78,54 \%& - \\
Branch Coverage & 89,0 \% & 58,2 \% & 59,36 \%& - \\
\hline
$f_4()$ (19 lines, 4 branches )& & \multicolumn{2}{c|}{} & \\
\hline
Time & 0.062s & \multicolumn{2}{c|}{0.040s}  & $<$ 1s \\
Nr of Test Cases & 3 & \multicolumn{2}{c|}{3} & 9 \\
Lines Coverage &  100 \% & \multicolumn{2}{c|}{100 \%} & 100 \% \\
Branch Coverage & 100 \% & \multicolumn{2}{c|}{100 \%}& 100 \% \\
\hline
$f_5()$ (28 lines, 35 branches )& & \multicolumn{2}{c|}{}  & \\
\hline
Time & 0.337s & 3.234s & 41,176s & 10s \\
Nr of Test Cases & 3 & 463 & 2187 & 201 \\
Lines Coverage &  100 \% & 100 \% & 100 \%& 85,71 \% \\
Branch Coverage & 100 \% & 100 \% & 100 \%& 85,71 \% \\
\hline
\end{tabular} 

\vspace{10mm}

\begin{tabular}{l  r }
\begin{tabular}{|r | c | c | c |}
\hline
  & CTGEN & KLEE & PathCrawler \\
\hline
$Tritype()$ & & &  \\
\hline
Time & 8.404 & 0.095  & $<$ 1s \\
Nr of Test Cases & 8 & 1 & 11 \\
Lines Coverage &  100 \% & 41,66 \%  & 100 \% \\
Branch Coverage & 100 \% & 20 \%  & 100 \% \\
\hline
\end{tabular} &
\begin{lstlisting}
int Tritype(double i, double j, double k){
  int trityp = 0;
  if (i < 0.0 || j < 0.0 || k < 0.0)          
    return 3;
  if (i + j <= k || j + k <= i || k + i <= j) 
    return 3;    
  if (i == j) trityp = trityp + 1;            
  if (i == k) trityp = trityp + 1;            
  if (j == k) trityp = trityp + 1;            
  if (trityp >= 2)                            
      trityp = 2;
  return trityp;
}
\end{lstlisting}
\\
\end{tabular}

\vspace{10mm}

\begin{tabular}{l  r }
\begin{tabular}{|r | c | c | c |}
\hline
  & CTGEN & KLEE & PathCrawler \\
\hline
$alloc\_ptr()$ & & &  \\
\hline
Time & 0.071s & 0.064s  & $<$ 1s \\
Nr of Test Cases & 4 & 4 & 2 \\
Lines Coverage &  100 \% & 100 \%  & 42,86 \% \\
Branch Coverage & 100 \% & 100 \%  & 50 \% \\
\hline
\end{tabular} &
\begin{lstlisting}
char *alloc_ptr(char *allocbufp, char *allocp, 
                unsigned int n)
{
    if(allocbufp == 0 || allocp == 0)
	return 0;
    
    if(allocbufp + ALLOCSIZE - allocp >= n){
	allocp += n;
	return allocp - n;
    } 
    return 0;
}
\end{lstlisting}
\\
\end{tabular}

\vspace{10mm}

\begin{tabular}{l  r }
\begin{tabular}{|r | c | c | c |}
\hline
  & CTGEN & KLEE & PathCrawler \\
\hline
$comp\_ptr()$ & & &  \\
\hline
Time & 0.032s & 0.055s  & $<$ 1s \\
Nr of Test Cases & 4 & 4 & 2 \\
Lines Coverage &  100 \% & 100 \%  & 75 \% \\
Branch Coverage & 100 \% & 100 \%  & 50 \% \\
\hline
\end{tabular} &
\begin{lstlisting}
int comp_ptr(char *p1, char *p2)
{
    if(p1 != NULL && p2 != NULL && p1 == p2){
	return 1;
    }
    return 0;
}
\end{lstlisting}
\\
\end{tabular}

\caption{Experimental results compared with other tools.}
\label{tab:results_comparison}
\normalsize
\end{table}


In comparison (see Table~\ref{tab:results_comparison}), experiments with  KLEE \cite{klee} and
PathCrawler \cite{pathcrawler:experience} demonstrated that CTGEN delivers
competitive results and   outperformed the others for the most complex function $f_1()$. The
experiments with PathCrawler were made with the online version
\cite{pathcrawler:online}, so it was not possible to exactly measure
the time spent by this tool. This tool, however, could not handle the complexity of $f_1()$, and KLEE 
did not achieve as much coverage as CTGEN, we assume that this is due to the path-coverage oriented  search strategy
which has not been optimised for achieving C1 coverage.

Functions  $f_4()$ and $f_5()$ are also taken from the
automotive testing project. 
Function $f_5()$ has  struct-inputs with bit fields. KLEE achieved 100 \% path
coverage. PathCrawler also targets path coverage, but due to
limitations of the online version (number of generated test cases,
available amount of memory) delivers only 201 test cases. However, we 
assume that without these limitations it will also
achieve 100 \% path coverage,  although in a larger amount of time than KLEE.
For the example function $Tritype()$  KLEE delivers bad results because it does not support
floating types; there is, however,   an extension KLEE-FP \cite{klee:fp}
targeting this problem. PathCrawler excels CTGEN and KLEE, but   can handle only the
\verb+double+ type, not \verb+float+, while CTGEN can calculate bit-precise solutions for 
both.
$alloc\_ptr()$ and $comp\_ptr()$ demonstrate handling of symbolic
pointers which is not supported by   PathCrawler; 
KLEE and CTGEN deliver comparable results.


\section{Related Work}\label{related_work}

The idea of using  symbolic  execution for test data generation
is not new, it is an active area of research since the 70's \cite{king,
  clarke}. In the past a number  of test data generation tools \cite{klee,
exe, pathcrawler:experience, bitblaze, klover, cbmc_indust, pex, cute,
dart, sage, pathfinder} were
introduced. Nevertheless to the best of our knowledge, only Pex (with Moles) supports automatic stub
generation as provided by CTGEN. Furthermore, CTGEN seems to be the only tool supporting traceability between test cases and requirements. From the experimental results available from other tool evaluations
we conclude that CTGEN outperforms most of them with respect to the UUT size that still can be handled for 
C1 coverage generation.

DART \cite{dart} is one of the first concolic testing tools to  generate
test data for C programs. It falls back to concrete values by external 
function calls, and does not support symbolic pointers.
CUTE \cite{cute} is also a concolic test data generator for C, and, 
like DART, falls back to concrete values  by
external function calls. It supports pointers, but collects only 
equalities/inequalities over them, while CTGEN   supports all
regular pointer arithmetic operations. 

SAGE \cite{sage} is built on DART, is a very powerful concolic testing
tool utilising whitebox fuzzing. It is fully automated and is in daily
use by Windows in software 
development process, and, accordingly to authors, uncovered about half of
all  bugs found in Windows 7. SAGE has a precise memory model, that
allows accurate pointer reasoning \cite{sage:pointers} and is very
effective because it works on large applications and not small
units, which allows to detect problems across components. Nevertheless
SAGE uses concrete values for sub-function calls which 
cannot be symbolically represented and, to our best knowledge, does not support 
 pre- and post-conditions.

Pex \cite{pex} is an automatic white-box test generation tool for .NET,
developed at Microsoft Research. It generates high coverage test
suites applying dynamic symbolic execution for parameterised unit tests
(PUT). Similarly to CTGEN it uses
annotations to define expected results, and the Z3 SMT Solver   to
decide on feasibility of execution paths. It also supports complex
pointer structures \cite{pex:pointers}. 
Pex with
Moles also offers a possibility to handle external function calls with
stubs; however, it is required, that the mock object is first
generated by the user with Moles to enable Pex to handle such calls, while CTGEN does it
automatically. 


Another approach using symbolic execution is applied by KLEE
\cite{klee}, the successor of EXE \cite{exe}. 
KLEE focuses on the interactions of the UUT with the running environment --
command-line arguments, files, environment variables etc. It redirects
calls accessing the environment to {\it models} describing external functions
in sufficient depth to allow generation of the path constraints required. 
Therefore  KLEE
can handle library  
functions symbolically only if a corresponding model  exists, and
all unmodelled library and external function calls are 
executed with concrete values. This may   reduce the  
coverage to be generated, due to random testing limitations. 
Furthermore KLEE does not
provide fully automated detection of inputs: they must be determined
by the user either by code instrumentation or by the command line
argument defining the number, size and types of symbolic
inputs.

Pathcrawler \cite{pathcrawler:experience} is another concolic testing tool. 
It tries to generate path coverage  for C functions.
In contrast to CTGEN, it  only supports one dimensional arrays and does not
support pointer comparisons and external function calls.
Pathcrawler proposed an approach, similar to ours for handling of
external function calls. Though, like Pex, it requires
mock objects to be defined first by the user, together with a formal
specification of its intended behaviour.

Another approach to test data generation
in productive industrial environments is based on bounded model checking  
\cite{cbmc_indust}. The authors used CBMC \cite{cbmc},  a  Bounded Model
Checker for ANSI-C and C++ programs, 
for the generation of test vectors.  The tool  supports pointer
dereferencing, arithmetic, dynamic memory and more. 
However,
since CBMC is applied here to generate a test case for each block of
the CFG of the UUT, CTGEN will be able to
achieve full decision coverage with fewer test cases in most situations. 
For
handling   external function calls the authors of  \cite{cbmc_indust} use nondeterministic
choice functions available in CBCM as stubs, and CBCM evaluates all traces
resulting from all possible choices. However, the tool can only simulate 
return values of external functions and does not consider the
possibility of manipulating values of global variables.
Although CBMC allows assertions and assumptions in the function body,
the authors  use them only to achieve the
branch coverage, not for checking functional properties.

PathFinder \cite{pathfinder} is a symbolic execution framework, that
uses a model checker to generate and explore different 
execution paths. PathFinder works on Java byte code, one of its main
applications is the production of test 
data for achieving high code coverage. PathFinder does not address pointer 
problems since these do not exist in Java. For handling of external function
calls the authors propose {\it mixed concrete-symbolic solving}
\cite{pathfinder:external}, which 
is more precise than CTGEN's solution with stubs - it will not generate
test data that is impossible in practise, but is incomplete,
i.e. there exist feasible paths, for which mixed concrete-symbolic
solving fails to find a solution. Furthermore, by definition of
accurate pre- and post-conditions the problem with impossible inputs
can be avoided by CTGEN.

Table~\ref{tab:checklist} summarises the results of our comparison.

\begin{table}[h]
\scalebox{0.65}{
\begin{tabular}{| p{3cm} | c | c | c | c | c | c | c | c | p{1,7cm}|}
\hline 
 &  CTGEN & PEX & CUTE & KLEE & PathCrawler & CBMC for SCS  & DART &
SAGE & PathFinder\\
\hline 
Platform & Linux & Windows & Linux & Linux & Linux &  & Linux & Windows &
Linux\\
Language & C & .NET & C & C & C & C & C & machine code & Java\\
\hline 
\hline
{\bf CAPABILITIES}  & & & & & & & & &\\
\hline 
C0 & Y & Y & Y & Y & Y & Y & Y & Y & Y\\
C1 & Y & Y & Y & Y & Y & Y & Y & Y & Y\\
MC/DC & N & Y & N & N & N & N & N & N & Y\\
C2 & N &  & Y & Y & Y & N & Y & Y & Y\\
\hline  
Pre-/Post & Y & Y & Y & Y & Y & N & N & N & Y\\
Requirements tracing & Y & N & N & N & N & N & N & N & N\\
Auxiliary vars & Y & N & NA & N & N & N & N & N & N\\
\hline 
Pointer arithmetics & Y & Y & N & Y & Y & Y & N & Y & -\\
Pointer dereferencing & Y & Y & N & Y & Y & Y & N & Y & -\\
Pointer comparison & Y & Y & Y & Y & N & Y & N & Y & -\\
Function pointer & N & NA & N & NA & N & Y & N & NA &\\
Arrays & Y & Y & Y & Y & P & Y &  & Y & Y\\
Symbolic offset & Y & NA & N & Y &  &  & N & Y &\\
Complex dynamic data structures (lists...) & N & Y & Y &  & N & Y & 
N &  & Y\\
\hline 
External function calls & Y & P & N & P & N & Y & N & N & P\\
Automatic stub handling & Y & P & N & N & N & N & N & N & N\\
Float/double & Y & N & N & N & Y & N & N & Y & Y\\
Recursion & N & Y & NA & Y & N & Y &  &  & Y\\
Multithreading & N & N & Y & N & N & N & N &  & Y\\
Automatic detection of inputs& Y & Y & N & N & Y & Y & Y & Y &\\
\hline
\hline 
{\bf TECHNIQUES} & & & & & & & & &\\
SMT solver & SONOLAR & Z3 & lp\_solver & STP & COLIBRI &  &
lp\_solver & Z3 & choco, IASolver, CVC3 \\
Concolic testing & N & Y & Y & Y & Y & N & Y & Y & N\\
STCT or acyclic graph with reNuse of nodes & STCT & STCT &  &
application states&  & 
transition relation &  &  & Y\\
Depth-first search & N & N & Y & N & Y & N & Y & N & N\\
\hline
\end{tabular}
}
\caption{Test Data Generating Tools.}
\label{tab:checklist}
\end{table}

\section{Conclusion}\label{conclusion}

In this paper we presented CTGEN, a  unit test generator for C functions.
CTGEN aims at test data generation for C0 or C1 coverage and supports test specifications with pre- and post-conditions and other annotation techniques, automated stub generation, as well as structure, pointer and array handling in the UUT.
The tool is targeted at embedded systems software testing. Its fitness for industrial scale 
testing campaigns has been demonstrated in a project from the automotive domain, 
where it effectively reduced   time and effort spent during  the testing phase.
Further experiments with other test data
generation tools showed that CTGEN produced competitive results and
in some aspects   outperformed the others.

To further improve performance and scalability of the tool we plan
to replace the current test case tree structure used for investigation of
function paths to be covered by  acyclic graphs allowing the  re-use of nodes.
The test generation technique is currently extended to support MC/DC coverage.

%
%

\bibliographystyle{eptcs}
\bibliography{ctgen}

\end{document}